\documentclass[sigconf]{acmart}

\renewcommand\footnotetextcopyrightpermission[1]{} 
\setcopyright{none}

\acmConference[SBES 2026]{40th Brazilian Symposium on Software Engineering}{September 8–11, 2026}{São Paulo, SP, Brazil}

\AtBeginDocument{
    
}

\usepackage{multirow}
\usepackage{multicol}
\usepackage{tcolorbox}
\definecolor{minhacor}{RGB}{220,220,220} 
\usepackage[table,xcdraw]{xcolor}
\usepackage{colortbl} 
\usepackage{enumitem}

\usepackage{subcaption}

\begin{document}

\title{Comprehending Python Repetition Structures: An Eye-Tracking Study with Novice Developers}
 
\author{José Júnior Silva da Costa}
\affiliation{
  \institution{Federal University of Campina Grande}
  \city{Campina Grande}
  \country{Brazil}
}
\email{josejunior@copin.ufcg.edu.br}

\author{Rohit Gheyi}
\affiliation{
  \institution{Federal University of Campina Grande}
  \city{Campina Grande}
  \country{Brazil}
}
\email{rohit@dsc.ufcg.edu.br}

\author{José Aldo Silva da Costa}
\affiliation{
  \institution{State University of Paraíba}
  \city{Patos}
  \country{Brazil}
}
\email{jose.aldo@servidor.uepb.edu.br}

\author{Márcio Ribeiro}
\affiliation{
  \institution{Federal University of Alagoas}
  \city{Alagoas}
  \country{Brazil}
}
\email{marcio@ic.ufal.br}

\renewcommand{\shortauthors}{Silva da Costa et al.}
\renewcommand{\shorttitle}{Comprehending Python Repetition Structures}

\begin{abstract}
Code comprehension is central to software development, maintenance, and evolution. In Python, the same computation can often be implemented using different repetition structures, such as \texttt{for} loops, \texttt{while} loops, recursion, and list comprehensions (LCs). These choices may affect readability, code review, onboarding, refactoring, and maintainability.
Prior studies have mostly compared repetition structures using correctness, completion time, error rates, and preferences, or have studied LCs in repositories in terms of prevalence, complexity, evolution, fault-proneness, and performance. Such evidence indicates which constructs are difficult, but provides limited insight into how developers visually process them and where comprehension effort concentrates.
In this paper, we conducted a controlled eye-tracking experiment with 32 undergraduate students with prior Python experience. Participants solved six Python comprehension tasks involving \texttt{for} loops, \texttt{while} loops, recursion, and LCs. Using a Latin Square design, we collected time, number of answer attempts, fixation duration, fixation count, and regression count over full code snippets and construct-specific \textit{Areas of Interest} (AOIs).
\texttt{for} loops were associated with the lowest visual effort in our tasks. Compared with \texttt{for}, \texttt{while} increased AOI fixation duration by up to 97\% and AOI regression count by 114\%, with regressions concentrated around counter-management statements. Recursion showed a descriptive 50\% increase in regression count, mainly through repeated transitions between the base case and the recursive call. LC increased AOI time by 62.5\% and AOI fixation duration by 80.9\%, while producing horizontal regression patterns consistent with token-by-token parsing. LC comparisons showed the clearest statistically significant pairwise differences, and the combined comparison across all non-\texttt{for} structures was significant for all eye-tracking metrics.
 {The results suggest that Python repetition structures may be associated with distinct visual-effort patterns:} \texttt{while} loops require manual state tracking, recursion requires call-flow reconstruction, and LCs require dense local parsing. These findings complement traditional performance metrics and provide process-level evidence for Python readability, code review, refactoring, onboarding, and maintainable-code guidelines.
\end{abstract}


\frenchspacing
\maketitle

\section{Introduction}

Code comprehension is a central activity in software engineering. Developers must understand existing code to fix defects, review changes, add features, refactor implementations, and onboard into software projects~\cite{brooks1978,xia2017measuring}.
Because comprehension often dominates software maintenance and evolution tasks, even small differences in how code is structured may affect developer productivity, code review effectiveness, and software
maintainability. In these activities, developers often inspect unfamiliar implementations and decide whether code is clear enough to maintain, review, or refactor.

In Python software projects, the same computation can often be implemented using different repetition structures, such as \texttt{for} loops, \texttt{while} loops, recursion, and list comprehensions (LCs). These alternatives are not merely stylistic. They differ in how they expose control flow, state updates,
termination conditions, and data transformations. For example, \texttt{for} loops encapsulate bounded iteration in a single loop declaration, \texttt{while} loops require readers to coordinate initialization, condition checking, and state updates, recursion
requires reasoning about base cases and recursive calls, and LCs compress iteration, filtering, and data construction into a single expression. As a result, functionally equivalent implementations may
impose different comprehension demands on developers who need to inspect, review, debug, or modify the code. Thus, two implementations that compute the same output may still differ substantially in readability and maintenance effort.

These constructs are relevant in practice since they appear in real Python repositories. Empirical studies show that \texttt{for} loops are among the most frequently used Python language features~\cite{peng2021empirical}, while LCs have been observed at scale in open-source projects and studied with respect to their complexity, evolution, fault-proneness, and performance characteristics~\cite{belias2022python,zampetti2022empirical,zid2024study}. Comparing these constructs is important not only for understanding how
Python is taught, but also for understanding how less-experienced developers comprehend code they may encounter during review and evolution tasks. We use \texttt{for} loops as the baseline because they represent an idiomatic and frequent form of bounded iteration in Python, while \texttt{while} loops, recursion, and LCs expose distinct comprehension mechanisms: explicit state management, call-flow reconstruction, and compact expression parsing, respectively.

Previous work has compared iteration and recursion using
outcome measures such as task correctness, error rates, completion time, and student preferences
~\cite{endres2021analysis,sulov2016iteration,esteero2018recursion}. Other studies have investigated LCs in Python repositories, focusing on their prevalence, complexity, evolution, fault-proneness, or performance~\cite{belias2022python,zampetti2022empirical,zid2024study}. However, these studies leave an important gap for software engineering: they do not explain how developers visually process alternative implementations of the same computation. In maintenance and review
tasks, this distinction matters since comprehension failures are not uniform. A developer may struggle to track state across a \texttt{while} loop, to relate a recursive call to its base case, or to parse the compact syntax of an LC. Identifying these structure-specific
bottlenecks requires observing the reading process itself. Without such process-level evidence, guidelines and refactoring recommendations may treat repetition structures as interchangeable stylistic alternatives, overlooking their different comprehension costs.

Eye tracking offers a way to investigate this process directly. By measuring fixation duration, fixation count, and regression count over both entire code snippets and specific \textit{Areas of Interest}
(AOIs), eye tracking can reveal where readers allocate attention and which parts of a construct require rereading. This fine-grained evidence is particularly useful for software engineering because it can
inform readability guidelines, code review practices, refactoring decisions, and recommendations for writing maintainable Python code. Rather than treating visual effort as a single global measure, AOI-level
analysis allows us to distinguish different comprehension bottlenecks, such as manual state tracking in \texttt{while} loops, call-flow reconstruction in recursion, and dense local parsing in LCs.

Despite this potential, few studies have systematically compared \texttt{for}, \texttt{while}, recursion, and LCs within the same experimental design using eye-tracking data. This gap limits our
understanding of how different Python repetition structures affect code comprehension and which syntactic or semantic elements contribute most to developers' visual effort. Consequently, we still lack evidence explaining not only whether some structures are harder to comprehend, but also where this additional effort concentrates during code reading.

In this paper, we conducted a controlled experiment with 32 undergraduate students with prior Python experience, 
 {whom we refer to as novice developers in early stages of professional formation. Our focus is on the cognitive mechanisms underlying how novices comprehend repetition structures, with implications for both computing education and industry practices such as code review, maintenance, and onboarding.}
Participants solved six Python comprehension tasks involving \texttt{for} loops, \texttt{while} loops, recursion, and LCs. We used a Latin Square design to control for task and ordering effects, and we
analyzed both task performance and visual effort. Eye-tracking metrics were computed over the whole code snippet and over AOIs corresponding to each repetition structure.

Our results show that \texttt{for} loops were associated with the lowest visual effort across the analyzed metrics. Compared with \texttt{for},
\texttt{while} loops increased fixation duration by up to 97\% and regression count by 114\% within the AOI, with regressions concentrated around counter-management statements. Recursion showed a descriptive 50\% increase in regression count, mainly due to repeated transitions between the base case and the recursive call. LCs produced the clearest statistically significant pairwise differences, increasing AOI time by 62.5\% and fixation duration by 80.9\%, while also producing horizontal regression patterns within a single line, consistent with token-by-token parsing. Although only LC time comparisons reached statistical significance when analyzed in isolation, the
combined comparison across all repetition structures was significant for all eye-tracking metrics. 

Overall, these results suggest that repetition structures differ not only in the amount of visual effort they require, but also in the type of comprehension effort they induce. This study contributes process-level evidence for how novice developers inspect alternative implementations of equivalent Python computations, informing readability decisions in code review, onboarding, maintenance, refactoring, and Python style guidelines. Rather than showing that one construct is universally preferable, our findings indicate that different constructs create different comprehension demands.

\section{Study Definition}
\label{sec:Study Definition}

Following the Goal-Question-Metrics approach~\cite{basili1994thegqm}, 
\textbf{we compare} the \texttt{for} loop, the \texttt{while} loop, recursion, and LC 
\textbf{with the purpose of} understanding how these repetition structures impact visual effort and code comprehension \textbf{from the point of view} of novice Python programmers \textbf{in the context of} tasks adapted from  introductory programming courses. We address the following research questions:
\begin{description}[leftmargin=0.5cm,labelsep=0.3cm,itemindent=0pt]
    \item[\textbf{RQ$_1$}] \textit{What is the impact of code structures on task-solving time?} We measure task time and AOI time for the full code and the repetition-structure \textit{Area of Interest} (AOI),
    as time is commonly used as a proxy for comprehension
    effort~\cite{sharif2010eye,da2021evaluating}.

    \item[\textbf{RQ$_2$}] \textit{What is the impact of code structures on the number of attempts?} We measure submission count from task start
    until a correct answer or abandonment, as an indicator of task
    difficulty and confusion~\cite{da2023seeing}.

    \item[\textbf{RQ$_3$}] \textit{What is the impact of code structures on fixation duration?} We measure total fixation duration over the full
    code and AOI, since longer fixations indicate higher attentional
    demands and cognitive load~\cite{crosby2002theroles}.

    \item[\textbf{RQ$_4$}] \textit{What is the impact of code structures on fixation count?} We measure the number of fixations over the full
    code and AOI, as more fixations indicate extended processing time
    and visual effort~\cite{sharafi2012women}.

    \item[\textbf{RQ$_5$}] \textit{What is the impact of code structures on regression count?} We measure backward eye movements over the full
    code and AOI, since regressions signal comprehension
    difficulty~\cite{busjahn2015eye,sharafi2015eye}.
\end{description}

\section{Methodology}
\label{sec:Methodology}

We describe the methodology used in this study.

\subsection{Pilot Study}\label{estudoPiloto}

{We conducted pilot studies in two rounds (six, then four additional participants) to refine the protocol; these data were excluded from the final analysis, which covers only the 32 main-experiment participants.}
To carry out the pilot studies and the {actual} experiment,{ we used code tasks,} a characterization form, a consent form, and a questionnaire for a semi-structured interview. {We evaluated the programs by testing code snippets with varying levels of difficulty. We standardized the code font size, font style, line spacing, and indentation. Additionally, we assessed the questions on the forms and the questionnaire}.

All participants were Brazilian and native speakers of Portuguese. We used the vocabulary of the programs in Brazilian Portuguese, thus avoiding obstacles in understanding the vocabulary of the programs. The names of the methods and variables were selected and discussed by the researchers. Names such as ``result'' were used to receive the results of operations. We sought to avoid names that made it too easy to carry out the operations, opting for more neutral names such as ``calculate'', with the aim of having the participant analyze the code.

We adjusted and refined the names of the methods and variables in the pilot studies, testing how well the names presented the intention of the methods. The names were discussed by the researchers to find the best options. The experiment was organized into five phases: (1) Characterization, (2) Tutorial, (3) Warm-up, (4) Tasks and (5) Qualitative interview. {The experiment was estimated to take around 60 minutes per participant.}

\subsection{Experiment Phases}\label{ExperimentPhases}

In the first phase, we explained the study and what data would be captured. At this stage, each participant filled out a consent form agreeing to participate and acknowledging that their identity would remain anonymous. They also filled out a characterization form with questions about their experience with programming.
In the second phase, we provided a tutorial explaining the experimental procedure. In this phase, participants received instructions about the eye-tracking camera and how to perform the tasks. After this, the eye tracker was calibrated for each participant. For calibration, the participant must look at the locations on the screen indicated by the camera software. At the end of the calibration, the camera software also reports when the calibration was successful.

In the third phase, each participant warms up for the experiment by solving a simple problem. During the warm-up, we demonstrated how participants should verbally provide the program output. Participants are also instructed to close their eyes for two seconds before and after solving the problem, and how it would be signaled whether the answer is correct or incorrect. After warming up, participants can become more comfortable with the setup of the experiment and the equipment used.
In the fourth phase, we {run the experiment in which each student solves six program tasks}. To avoid the learning effect, {we used} the Latin Square design~\cite{box2005statistics}, which  {is} explained in more detail in Section~\ref{Treatments}.
In the fifth phase, the experiment ends with a semi-structured interview. The interview explores participants' approach to the programs and their impressions. For each program, three questions were asked:

\begin{itemize}
    \item What strategy did you use to find the program output?
    \item How do you evaluate the difficulty of the task: very easy, easy, neutral, difficult or very difficult?
    \item What were the main difficulties involved, if any? Could you point them out in the program?
\end{itemize}

{These three questions were used to triangulate RQ$_{1}$–RQ$_{5}$ as a whole, connecting participants' self-reported strategies and difficulties with the quantitative and eye-tracking evidence discussed in Section \ref{sec:Results and Discussion}.}
We assessed participants’ preferences using a five-point scale by presenting two code snippets (A and B) and asking which they preferred, ranging from strong preference for A to strong preference for B, along with a justification for their choice.
To conduct the experiment, participants were seated in a fixed chair to improve eye-tracking accuracy. Due to camera limitations, perfect data capture was not always possible; we adjusted fixation blocks along the \textit{y}-axis during analysis. This procedure is discussed in Section~\ref{sec:Threats to Validity}.

\subsection{Subjects}\label{Subjects}

{For the pilot study, we recruited six undergraduates and later added four more after adjusting the design, totaling ten participants. They reported having between 6 and {72} months of experience with programming languages in general, including mainly Python, Java and C. Participants were recruited mainly in person from two universities in one Brazilian city. All were Portuguese-speaking undergraduate students. 

{For the main experiment, we evaluated 32 undergraduate students, whom we refer to as novices.} In some cases, participants had more experience because they had already been programming before starting their undergraduate course. Participants were recruited primarily in person from three universities in two Brazilian cities. All were Portuguese-speaking undergraduates from different academic semesters.}
 {The characterization of the 32 participants is provided in the study's supplementary artifact package~\cite{anonymous2026supplementary}.}

\subsection{Treatments}
\label{Treatments}

\begin{figure}[]
\centering
\includegraphics[width=5.5cm]{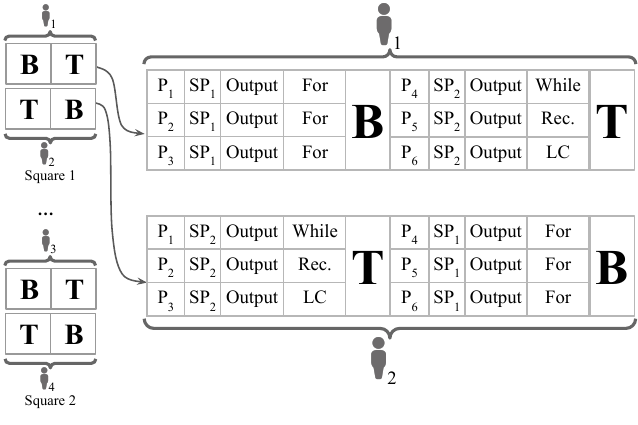}
\caption{Experiment structure. Programs (P) are distributed into Sets of Programs (SP), and each participant solves output-prediction tasks under baseline (B: \texttt{for} loops) and treatment (T: \texttt{while} loops, recursion, or LC) conditions.}
\label{fig: metodologia}
\end{figure}

As illustrated in Figure~\ref{fig: metodologia}, each participant analyzed six programs (P1--P6). Each program corresponded to an output-prediction task and had two alternative implementations: a baseline version using the \texttt{for} loop and a treatment version using either the \texttt{while} loop, recursion, or list comprehension (LC). To minimize learning effects, we adopted a Latin Square~\cite{box2005statistics} design to assign the programs, so that no participant solved the same task in both its baseline and treatment versions. Twelve different programs were designed and divided into two Sets of Programs (SP$_{1}$ and SP$_{2}$). Each participant analyzed three programs from SP$_{1}$ and three from SP$_{2}$, totaling six tasks. The programs within the same set were designed to produce different outputs for the given inputs. Across the Latin Squares, the sets alternated between baseline and treatment roles: when one set was presented with the \texttt{for} loop, the other was presented with the \texttt{while} loop, recursion, or LC, and vice versa. Thus, the \texttt{for} loop was always the baseline structure, whereas the \texttt{while} loop, recursion, and LC formed the treatment structures. In all programs, participants had to specify the correct output without multiple answer options. The tasks included standard problems such as calculating factorials, summing elements in a list, and similar operations.

\subsection{Programs}\label{programs}

We selected and adapted the programs through a manual analysis of repositories with introductory programming tasks. The main source of the programs was \textit{GeeksforGeeks}, which is popular for learning and practicing programming. For the experiment, tasks with small and complete programs were selected, summarized in Table~\ref{TasksDescription}. These code snippets were adapted considering the needs of the experiment and the limitations of the eye-tracking camera. The selected tasks enabled controlled comparisons between repetition structures while preserving equivalent task logic. By implementing the same computational problems with the \texttt{for} loop, the \texttt{while} loop, recursion, and list comprehension, we aimed to isolate the effect of the structure used to express repetition on comprehension and visual effort. Our analysis focuses on tasks in which the number of iterations is known and fixed in advance. These constructs are among the most fundamental in programming and frequently appear in introductory materials, making them appropriate for assessing how novices understand different ways of expressing iteration. For each task, the participant needed to provide the correct output in an open response format, i.e., no response options were provided. 

\begin{table}[]
\centering
\footnotesize
\caption{Tasks used in evaluation.}
\begin{tabular}{p{2cm}p{3.6cm}p{0.6cm}p{0.8cm}}
\toprule
\textbf{Task} & \textbf{Description} & \textbf{Base} & \textbf{Treat.} \\ 
\midrule
Sum list & 
Traverse a list and sum all elements & 
\texttt{for} & \texttt{while} \\
\midrule
Multiples of three & 
Count elements that are multiples of three & 
\texttt{for} & \texttt{while} \\
\midrule
Sum from one to n & 
Sum numbers from one up to a given number & 
\texttt{for} & rec. \\
\midrule
Exponentiation & 
Calculate value1 raised to the power of value2 & 
\texttt{for} & rec. \\
\midrule
Quantity of 0 in list & 
Count elements equal to zero & 
\texttt{for} & LC \\
\midrule
Count even numbers & 
Count even elements in a list & 
\texttt{for} & LC \\
\bottomrule
\end{tabular}
\label{TasksDescription}
\end{table}

The evaluated programs had between 4 and 8 lines of code. The number of lines was limited so that the programs would fit completely on the screen. The programs were checked to remove syntax errors. They were also organized according to the PEP 8 (Python Enhancement Proposal) standard~\cite{van2001pep8}, a style guide for Python code, using the Python Syntax Checker PEP 8 tool. The Consolas font style was used, size 12, 1.5 line spacing, and eight blank spaces for indentation. We also used simple constructs that commonly occur in many programming languages.

The \texttt{for} loop is commonly used to iterate over elements in iterable structures, allowing each element to be processed sequentially through the loop body. This loop structure is specifically designed for scenarios where the number of repetitions is predetermined, making it particularly useful when iterating over fixed lists of elements such as lists, tuples, dictionaries, or strings. The \texttt{for} loop uses values derived from iterable structures. The iterable serves as a source for a sequence of values, each linked to the loop variable. With each iteration, the set of instructions within the loop is executed after assigning the current value to the loop variable~\cite{bailey2013python}.

The \texttt{while} structure in Python is similar to other programming languages such as C and Java, where a condition is tested, and if true, the instructions in the following block are executed in a loop. These instructions potentially change the state of the condition, so the condition is tested again and the block may be executed again. This process continues until the condition becomes \texttt{False}; if the condition is initially \texttt{False}, the block is never executed. If the condition does not become \texttt{False}, the loop continues indefinitely~\cite{bailey2013python}.
Recursion involves a function calling itself repeatedly, executing a set of instructions until a stopping condition is satisfied.
List Comprehension (LC)~\cite{bailey2013python} allows generating \textit{tuple}, \textit{list}, \textit{dictionary}, and \textit{set} objects from loops over iterable structures.

We use the \texttt{for} loop as the baseline, not because the other constructs are less relevant, but because the \texttt{for} loop represents an idiomatic and frequently used form of bounded iteration in Python~\cite{peng2021empirical}. The treatment constructs expose different comprehension mechanisms: \texttt{while} loops make loop control explicit through conditions and state updates, recursion requires reasoning about base cases and recursive calls, and LCs compress iteration and data construction into a single expression. Thus, our goal is not to compare constructs only by how often they occur in repositories, but to understand how different ways of expressing repetition affect novice developers' visual effort when solving equivalent programming tasks.

\subsection{Eye-Tracking System}\label{eyeTrackingSystem}

For the experiment, the Tobii Eye Tracker 4C equipment was used, which has a sample rate of 90 Hz. Eye-tracking calibration followed the device driver's standard procedure with five points. The eye-tracking camera was mounted on a laptop screen with a resolution of 1366 x 768 pixels, height of 17.4 cm and width of 30.9 cm, at a distance of 50-60 cm from the participant. Each task was presented in an image in full-screen mode, but an Integrated Development Environment (IDE) was not used, nor was the {line numbering}. An accuracy error of 0.7 degrees was calculated, which translates into 0.6 lines of printing on the screen, considering the font size and line spacing. The line {spacing was} designed to be large enough to overcome the accuracy limitations of the eye tracker.  {To process and analyze the data, we used a Python script developed by the authors. We include the script, raw and processed eye-tracking data, in the study's supplementary material.}

\subsection{Fixation and Saccade Identification}\label{FixationAndSaccadesInstrumentation}

We can understand fixations as pauses over certain regions of information of interest and saccades as movements performed quickly between fixations~\cite{xia2017measuring}. {In a fixation, attention rests on a region of a visual stimulus (e.g., a piece of source code), triggering cognitive processing}~\cite{sharafi2020practical}. The main eye-tracking metrics~\cite{sharafi2015eye} used in this work are briefly explained next.

\textit{Duration of fixations:} While examining a scene, our eyes maintain stability for a certain duration, enabling us to concentrate on specific elements. As soon as we see a word, for instance, we try to interpret it, directing our attention toward it until we understand it~\cite{just1980theory}. The position and duration of fixations have been associated with the focus of attention~\cite{crosby2002theroles}. Longer fixations have been associated with an increase in demands of attentiveness~\cite{busjahn2011analysis}. 

\textit{Number of fixations:} It refers to our ability to fixate our eyes on different locations to examine a scene. An increased number of fixations is indicative of an extended processing time required to comprehend code as well as more attention to the code~\cite{binkley2013impact,crosby2002theroles}. A high number of fixations is associated with a greater visual effort to answer a question~\cite{sharafi2012women}.

\textit{Eye-movement regressions:} These regressions consist of backward eye movements over the stimulus~\cite{busjahn2015eye}. In code comprehension tasks, they correspond to visually returning to earlier parts of the code, either within the same line or across previous lines~\cite{sharafi2015eye}. 

{Gaze samples are classified into fixations based on a duration threshold, which varies across the literature depending on task type: 200–300 ms~\cite{rayner1998eye} or 100–200 ms~\cite{salvucci2000identifying}. Following prior code-comprehension studies~\cite{da2021evaluating,da2023seeing}, we used 200 ms.}
To classify fixations, we used Dispersion-Threshold Identification (I-DT)~\cite{salvucci2000identifying} and classified gaze samples as belonging to a fixation if they were located within a spatial region of approximately 0.5 degrees~\cite{nystrom2010adaptive}. On our screen it corresponded to 25 pixels.

\subsection{Analysis of the Results}
\label{AnalysisResults}

The experiment involved 32 participants. There were 192 programs in total, of which the subjects solved 143, corresponding to 74.5\% of the total. Among the 49 unsolved or abandoned tasks, 21 occurred in the Baseline group, and 28 in the Treatment group. In general, the subjects used between one and three attempts to solve or give up. {The eye tracker showed calibration limitations, causing systematic gaze misalignment. We detected this in longer-line programs by inspecting heatmaps and fixation plots for gaze points falling on white space instead of code. Following~\cite{da2023seeing}, we uniformly corrected all fixations for programs with such consistent offsets.}

{We defined AOIs for each task (e.g., list definitions, variable definitions, conditional statements) with a 10-pixel margin on all sides to avoid overlap given the camera's limitations. White space was also treated as a region, given the same calibration constraints. By analyzing the chronological order and position of fixations, we identified the sequence of regions visited by each participant.}

For the statistical analysis, we first assessed the normality of each metric using the Shapiro--Wilk test~\cite{shapiro1965analysis}, following guidelines for eye-tracking studies~\cite{sharafi2020practical,da2021evaluating,da2023seeing}. Since the data did not consistently follow a normal distribution, we used non-parametric tests for the comparisons between baseline and treatment conditions. Specifically, we used the Wilcoxon rank-sum test, also known as the Mann--Whitney U test, as implemented by the \texttt{wilcox.test} function in R. We used this test because participants did not solve the same program in both its baseline and treatment versions; therefore, the observations compared between conditions were treated as independent. We adopted a significance level of $\alpha = 0.05$.

We report the percentage difference (PD) to quantify the relative change between treatment and baseline conditions. For median-based metrics, PD was computed as $(M_T - M_B) / M_B \times 100$, where $M_T$ and $M_B$ denote the treatment and baseline medians, respectively. For the number of attempts, we used the mean instead of the median because the values varied little across tasks and the mean is more sensitive to small differences. Positive PD values indicate higher values for the treatment condition.

We did not apply a formal correction for multiple comparisons because the analysis aims to characterize patterns across complementary comprehension metrics rather than to draw conclusions from isolated tests. To reduce the risk of overinterpretation, we interpret the results by considering the consistency across quantitative metrics, eye-tracking visualizations, and participants' qualitative feedback.
 {Nonetheless, to assess the robustness of our findings under a more conservative threshold, we performed a sensitivity analysis applying Bonferroni correction within each family of comparisons. Time-related effects for code-level metrics remained robust (For vs. LC and All), while the remaining metrics no longer reached significance under this stricter threshold, consistent with the modest sample sizes typical of eye-tracking studies.}

\section{Results}
\label{AnsweringResearchQuestions}

This section presents the results for each research question. Table~\ref{tab:allresults} summarizes the main results of our work. 

\begin{table*}[]
\centering
\footnotesize
\setlength{\tabcolsep}{4pt}
\caption{Comparative analysis across all metrics. PD denotes the percentage difference between the treatment structure and the \texttt{for} baseline; positive values indicate higher values for the treatment condition. \textit{p} denotes the \textit{p}-value obtained with the Wilcoxon rank-sum test after assessing normality with the Shapiro--Wilk test. ES denotes the effect size, reported for statistically significant comparisons; -- indicates non-significant comparisons for which ES is not shown. Bold \textit{p} values indicate statistically significant differences at $p \leq 0.05$. {The `All' row, aggregating the three treatment structures (while, recursion, LC), is a complementary summary, not a replacement for the individual treatment comparisons.}}
\label{tab:allresults}
\begin{tabular}{llrrrrrrrrrrrrrrr}
\toprule
\multirow{2}{*}{\textbf{Comp.}} & 
\multirow{2}{*}{\textbf{Ctx.}} &
\multicolumn{3}{c}{\textbf{Time (s)}} &
\multicolumn{3}{c}{\textbf{Attempts}} &
\multicolumn{3}{c}{\textbf{Fix. Dur. (s)}} &
\multicolumn{3}{c}{\textbf{Fix. Count}} &
\multicolumn{3}{c}{\textbf{Regressions}} \\
\cmidrule(lr){3-5}\cmidrule(lr){6-8}\cmidrule(lr){9-11}
\cmidrule(lr){12-14}\cmidrule(lr){15-17}
& & PD & \textit{p} & ES & PD & \textit{p} & ES & PD & \textit{p} & ES & PD & \textit{p} & ES & PD & \textit{p} & ES \\
\midrule

\multirow{2}{*}{For vs. While}
& Code & ↑59.54 & 0.30 & -- 
& \multirow{2}{*}{↑12} & \multirow{2}{*}{0.06} & \multirow{2}{*}{--}
& ↑47.02 & 0.19 & -- & ↑59.49 & 0.21 & -- & ↑48.57 & 0.19 & -- \\
& AOI  & ↑95.00 & 0.13 & -- & & &
& ↑97.22 & 0.13 & -- & ↑95.45 & 0.12 & -- & ↑114.29 & 0.15 & -- \\
\midrule

\multirow{2}{*}{For vs. Rec.}
& Code & ↑13.10 & 0.31 & --
& \multirow{2}{*}{↑3.4} & \multirow{2}{*}{0.75} & \multirow{2}{*}{--}
& ↑15.93 & 0.33 & -- & ↑25.97 & 0.28 & -- & ↑50.00 & 0.21 & -- \\
& AOI  & ↑13.49 & 0.57 & -- & & &
& ↑29.76 & 0.43 & -- & ↑22.86 & 0.34 & -- & ↑23.08 & 0.35 & -- \\
\midrule

\multirow{2}{*}{For vs. LC}
& Code & ↑55.20 & \textbf{0.01} & –0.55
& \multirow{2}{*}{↑24.5} & \multirow{2}{*}{0.06} & \multirow{2}{*}{--}
& ↑51.00 & 0.06 & -- & ↑57.14 & 0.06 & -- & ↑90.00 & 0.09 & -- \\
& AOI  & ↑62.50 & \textbf{0.02} & –0.50 & & &
& ↑80.96 & 0.06 & -- & ↑61.54 & 0.08 & -- & ↑100.00 & 0.10 & -- \\
\midrule

\multirow{2}{*}{All}
& Code & ↑80.50 & \textbf{0.01} & –0.36
& \multirow{2}{*}{↑15.79} & \multirow{2}{*}{\textbf{0.03}} & 
\multirow{2}{*}{0.46}
& ↑65.92 & \textbf{0.02} & –0.32 
& ↑65.79 & \textbf{0.02} & –0.46 
& ↑76.67 & \textbf{0.02} & –0.44 \\
& AOI  & ↑92.70 & \textbf{0.01} & 0.48 & & &
& ↑102.3 & \textbf{0.02} & –0.31 
& ↑100.00 & \textbf{0.02} & –0.43 
& ↑94.44 & \textbf{0.03} & –0.30 \\
\bottomrule
\end{tabular}
\end{table*}

\subsection{RQ$_1$: Time}\label{RQTime}

We use medians for time-based metrics because they are less sensitive to extreme values. In the \textit{for vs. while} comparison, the \texttt{while} loop resulted in a 59.54\% increase in time spent on the code and a 95.0\% increase in time spent in the AOI compared to the \texttt{for} loop. These differences were not statistically significant (\textit{p}=0.30 for code and \textit{p}=0.13 for AOI).
In the \textit{for vs. recursion} comparison, recursion tasks showed smaller increases: 13.10\% for code time and 13.49\% for AOI time. These differences were also not statistically significant (\textit{p}=0.31 and \textit{p}=0.57, respectively).
The clearest pairwise result was observed for \textit{for vs. LC}. LC tasks increased code time by 55.20\% and AOI time by 62.50\%, with statistically significant differences in both contexts (\textit{p}=0.01 and \textit{p}=0.02, respectively). When all treatment structures were analyzed together against the \texttt{for} loop, time also increased significantly in both the code and AOI contexts.
RQ$_1$ indicates that \texttt{for} loops were associated with lower task-solving time in our tasks. However, among the pairwise comparisons, only LC showed statistically significant differences for time.

\subsection{RQ$_2$: Number of Attempts}\label{RQAttempts}

For the number of attempts, we use the average because the values vary little across tasks and the mean is more sensitive to small differences. Compared with the \texttt{for} loop, attempts increased by 12\% for the \texttt{while} loop, 3.4\% for recursion, and 24.5\% for LC. None of these pairwise comparisons reached statistical significance, although \texttt{while} and LC were close to the threshold (\textit{p}=0.06 in both cases).
When all treatment structures were analyzed together, the number of attempts increased by 15.79\% compared with the \texttt{for} loop, and this difference was statistically significant (\textit{p}=0.03). Thus, RQ$_2$ suggests that non-\texttt{for} structures tended to require more attempts overall, although the evidence is stronger for the aggregated comparison than for individual pairwise comparisons.

\subsection{RQ$_3$: Fixation Duration}
\label{RQFixationDuration}

Fixation duration captures the total time participants spent fixating on the code or AOI. We use medians for this metric because they are less sensitive to extreme values. Compared with \texttt{for}, fixation duration increased for all treatment structures. In the code context, the increases were 47.02\% for \texttt{while}, 15.93\% for recursion, and 51.00\% for LC. In the AOI context, the increases were 97.22\%, 29.76\%, and 80.96\%, respectively.
Despite these increases, none of the pairwise comparisons reached statistical significance. The LC comparison was the closest to significance in both contexts (\textit{p}=0.06 for code and \textit{p}=0.06 for AOI). In contrast, the aggregated comparison between all treatment structures and \texttt{for} was statistically significant for both code and AOI fixation duration (\textit{p}=0.02 in both cases).
Therefore, RQ$_3$ provides evidence that fixation duration was lower for \texttt{for} loops than for the other structures when analyzed collectively. At the pairwise level, the results should be interpreted as descriptive trends rather than conclusive statistical differences.

\subsection{RQ$_4$: Number of Fixations}
\label{RQNumberOfFixation}

The number of fixations also increased for all treatment structures compared with \texttt{for}. In the code context, the increases were 59.49\% for \texttt{while}, 25.97\% for recursion, and 57.14\% for LC. In the AOI context, the increases were 95.45\%, 22.86\%, and 61.54\%, respectively.
As with fixation duration, the pairwise comparisons did not reach statistical significance. The LC comparison again showed the closest values to the significance threshold, especially in the code context (\textit{p}=0.06). The aggregated comparison across all treatment structures was statistically significant for both code and AOI fixation count (\textit{p}=0.02 in both cases).
Thus, RQ$_4$ suggests that participants made more fixations when reading non-\texttt{for} structures overall, but the evidence for individual constructs is primarily descriptive.

\subsection{RQ$_5$: Number of Regressions}
\label{RQNumberOfRegressions}

Regression count increased across all treatment structures. In the code context, regressions increased by 48.57\% for \texttt{while}, 50.00\% for recursion, and 90.00\% for LC. In the AOI context, the increases were 114.29\%, 23.08\%, and 100.00\%, respectively. {Though not significant,} the LC comparison showed a relatively low p-value for code regressions (\textit{p}=0.09). When all treatment structures were grouped together, regressions increased significantly in both the code context (\textit{p}=0.02) and the AOI context (\textit{p}=0.03).
Overall, RQ$_5$ indicates that non-\texttt{for} structures were associated with more {regressions} when analyzed collectively.

\section{Discussion}
\label{sec:Results and Discussion}

{This section connects and discusses the quantitative results (Section~\ref{AnsweringResearchQuestions}) with eye-tracking visualizations and qualitative feedback.} 

\subsection{Comparing \texttt{for} and \texttt{while}
}\label{forVSwhile}

{Across all metrics, \texttt{while} showed higher values than \texttt{for}, increasing AOI time by 95\%, fixation count by 95.45\%, and regressions by 114.29\%. Though not statistically significant, the scarf plot helps explain this added visual effort.}

\begin{figure}[h!]
\centering
\includegraphics[width=9cm]{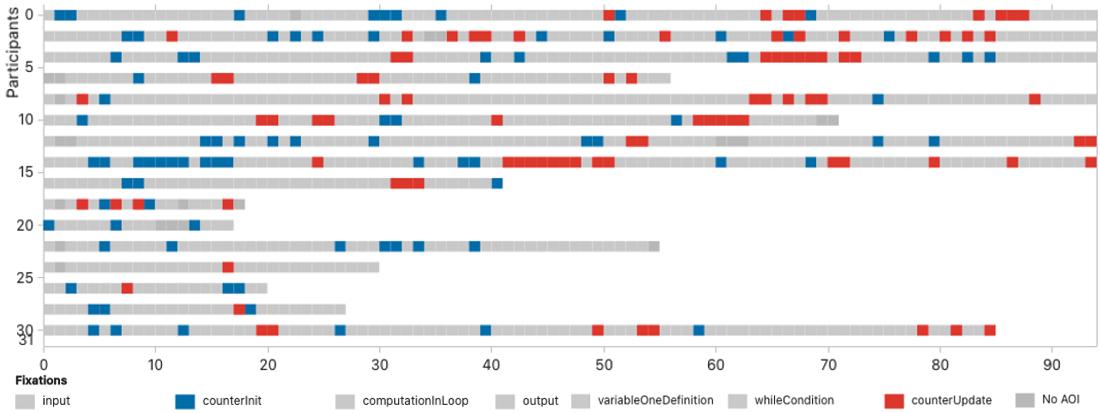}
\caption{Scarf plot for the \textit{Sum list} task, 
\texttt{while} version. Each row is one participant.}
\label{fig:scarfplot-forwhile}
\end{figure}

The scarf plot for the \texttt{while} version (Figure~\ref{fig:scarfplot-forwhile}) {suggests} two lines with no equivalent in the \texttt{for} version: \texttt{counterInit} (\texttt{counter = 0}) and \texttt{counterUpdate} (\texttt{counter = counter + 1}). These lines do not appear only at the beginning of the task. They recur throughout the entire reading session, visible as persistent purple and dark blue segments across nearly all participants.  {This recurrence is consistent with participants repeatedly returning} 
to verify the counter's initial value and increment mechanism, suggesting they lose track of loop state and must reconstruct it mentally during execution. These two lines represent a loop management cycle that runs in parallel with the problem-solving cycle (\texttt{whileCondition} $\rightarrow$ \texttt{computationInLoop}), competing for attentional resources throughout the task. In the \texttt{for} version, both responsibilities are absorbed into the loop declaration itself, \texttt{for element in range(0, len(lista))}, leaving no equivalent lines to track. The participant's attention is freed to focus entirely on the computation.  {Though non-significant, this pattern is consistent with additional effort from manual loop control.} 
The 114.29\% increase in AOI regressions for \texttt{while} is consistent with the repeated revisits to \texttt{counterInit} and \texttt{counterUpdate}, suggesting that manual counter management was the main source of the rereading.

\begin{figure}[t]
    \centering
    \begin{subfigure}[b]{0.23\textwidth}
        \centering
        \includegraphics[width=\textwidth]{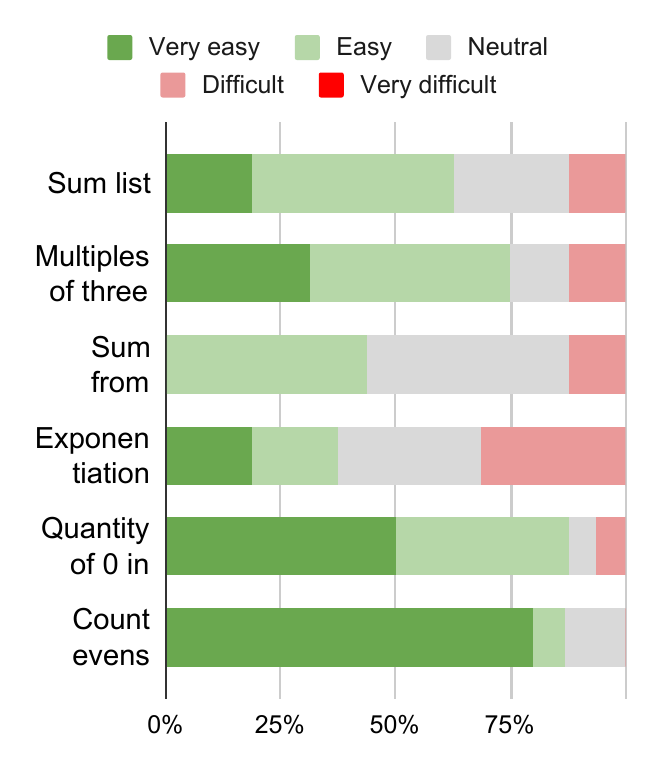}
        \caption{Baseline}
    \end{subfigure}
    \hfill
    \begin{subfigure}[b]{0.23\textwidth}
        \centering
        \includegraphics[width=\textwidth]{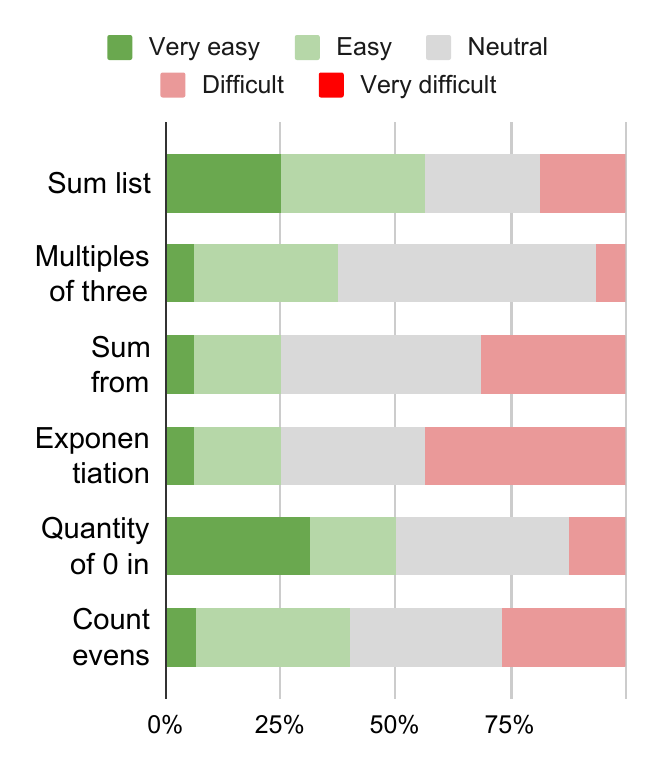}
        \caption{Treatment}
    \end{subfigure}
    \caption{Perception of task difficulty for different structures.}
    \label{fig: preferences Dissertacao}
\end{figure}

This is reflected in participants' own accounts: 
\textit{``while has all that counter stuff, I can get lost 
easier''} and \textit{``in \texttt{for} it already does this 
naturally''}. These comments provide qualitative support for the same pattern observed in the quantitative and visual data: participants did not simply spend more time on \texttt{while}; they repeatedly revisited the specific lines responsible for maintaining loop state.  {These self-reports corroborate especially RQ5 (regressions), linking the participants' own account of losing track of the counter to the repeated revisits on those lines, and are also consistent with the attempt and fixation patterns observed for RQ2–RQ4.} The eye-tracking data makes visible what students could only describe verbally. Difficulty ratings corroborate these findings: the \texttt{for} version was rated predominantly easy or neutral, while the \texttt{while} version shifted toward neutral and difficult 
(Figure~\ref{fig: preferences Dissertacao}).

This distribution of loop control across non-contiguous lines, namely, initialization, condition, and update, forces the reader to mentally integrate information spread throughout the code, rather than reading it as a unified construct. The \texttt{for} loop, by contrast, colocalizes these elements in a single declaration, enabling more integrated comprehension. For software engineering practice, this suggests that manual counter management may increase review and maintenance effort even in short code snippets, especially for less-experienced developers. For educators, this reinforces the value of introducing \texttt{for} as the default structure when iterations are known, reserving \texttt{while} for dynamic termination conditions.

\begin{figure}[]
\centering
\includegraphics[width=9cm]{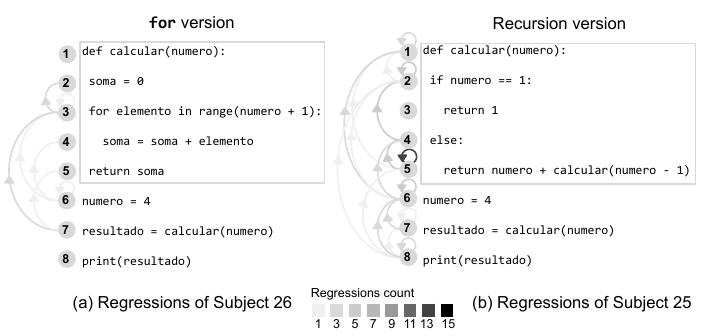}
\caption{Regression graph for two subjects examining code snippets, one with the \texttt{for} version (Subject 26) and one with the recursion version (Subject 25).  {This illustrates one representative pair; see Section~\ref{RQNumberOfRegressions} for the aggregated quantitative evidence across all recursion tasks.}}
\label{fig: Regressions25_21T_26_21B}
\end{figure}

\subsection{Comparing \texttt{for} and Recursion}
\label{forVSRecursion}

The quantitative results showed descriptive increases for recursion compared with \texttt{for}, including a 50\% increase in code regressions and a 23.08\% increase in AOI regressions. {Though non-significant}, the regression graph helps identify where this rereading effort occurred.
We observed higher effort for the tasks using recursion compared to \texttt{for}. Figure~\ref{fig: Regressions25_21T_26_21B} presents the regression graph comparing the \texttt{for} version (Subject 26) and the recursion version (Subject 25) for the \textit{Sum from one to n} task,  {illustrating a pattern that recurs across the
sample (see Section~\ref{RQNumberOfRegressions} for the aggregated analysis across all
recursion tasks). This concentration on the recursive call line was not confined to
Subject 25: Subjects 1, 9, 13, 15, and 31 each had 45.9--54.7\% of their individual
regressions directed at this single line (Line 5), reinforcing that the pattern generalizes beyond
this illustrative pair. Conversely, in the \texttt{for} condition, Subjects 6, 8, 12,
16, and 24 mirrored Subject 26's pattern, with 66.6--71.2\% of their regressions concentrated
on the loop lines (Lines 3 and 4).} For the \texttt{for} version, the subjects exhibited regressions indicating checks on how the loop affects the sum accumulation. In the recursion version, the subjects made many regressions suggesting difficulty in understanding how the recursive calls relate to the base case. The regression count was higher for the recursion version, reflecting greater cognitive load. This indicates that the additional regressions in the recursion version were not merely a general increase in rereading; they were associated with the need to connect the base case, the recursive call, and the returned value. 

\subsection{Comparing \texttt{for} and List Comprehension}\label{forVSLC}

\begin{figure*}[t]
\centering

\includegraphics[width=0.62\textwidth]{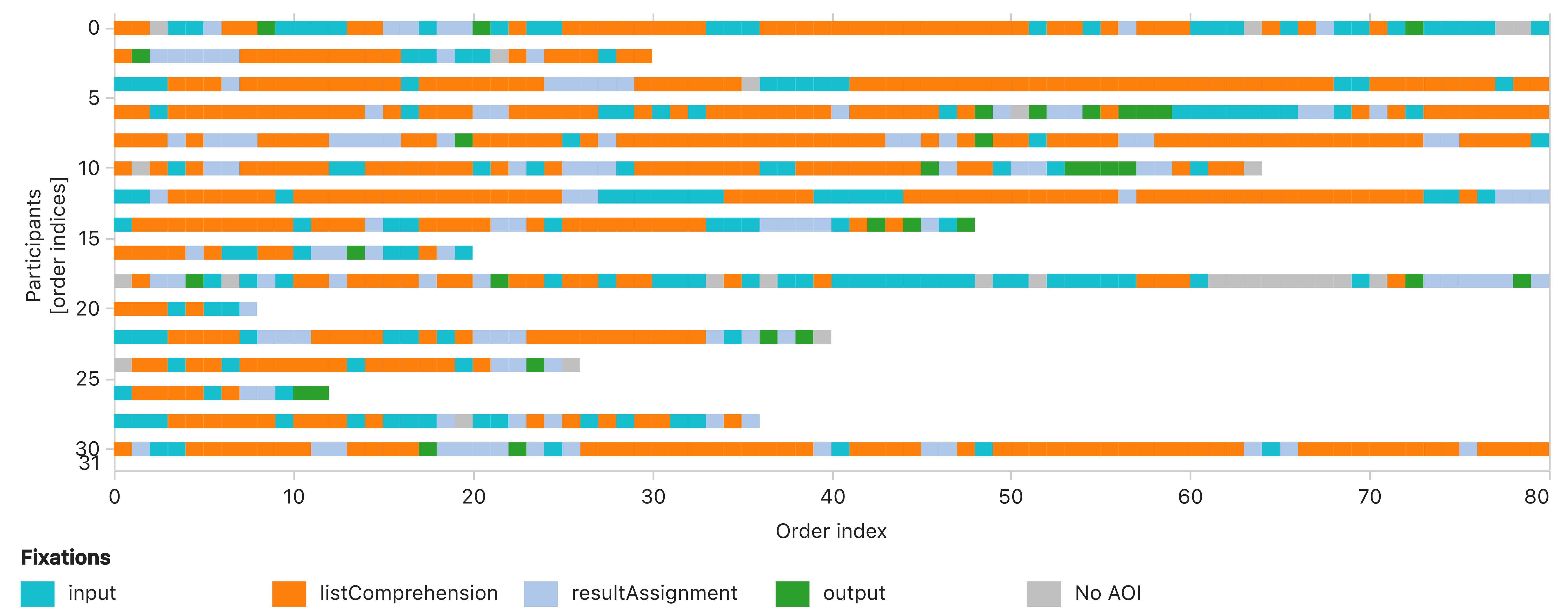}\\[-1mm]
{\small (a) List comprehension version}

\vspace{0.04cm}

\includegraphics[width=0.62\textwidth]{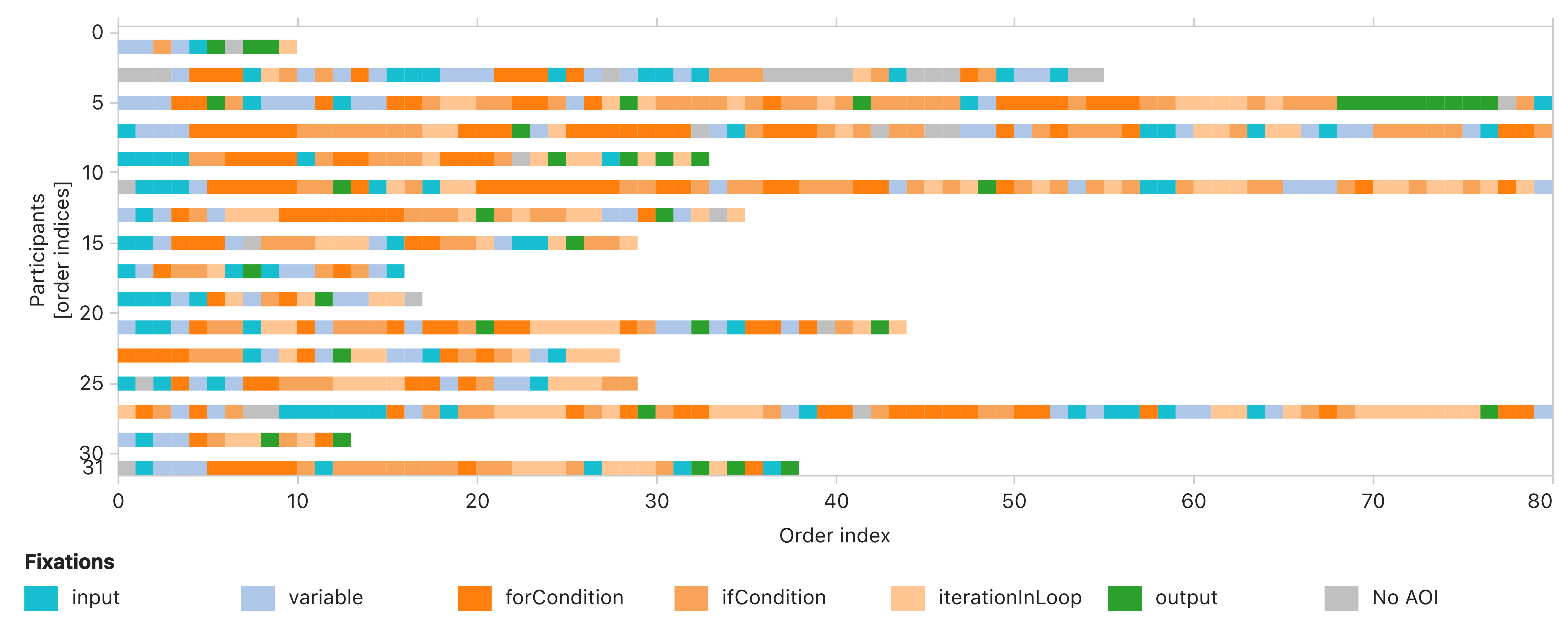}\\[-1mm]
{\small (b) \texttt{for} loop version}

\caption{
Scarf plots comparing (a) the LC and (b) the \texttt{for} implementations for the Quantity of zeros in the list task.
}
\label{fig:scarfplot}
\end{figure*}

In Section~\ref{AnsweringResearchQuestions}, we observed higher effort for LC tasks across all metrics. The LC comparison showed the clearest pairwise quantitative evidence. Compared with \texttt{for}, LC significantly increased code time by 55.20\% and AOI time by 62.50\%. It also showed large descriptive increases in fixation duration, fixation count, and regressions. Here we examine the underlying reading patterns that explain these differences.
Figure~\ref{fig:scarfplot} presents scarf plots for both versions of the task. Each row represents one participant, and each colored segment represents a fixation on a specific line of code, ordered sequentially from left to right. This visualization reveals the {reading trajectory} of each participant, which regions of the code were visited, in what order, and how often the eye returned to previously visited areas. The plots reveal a fundamental difference in reading behavior: in the \texttt{for} loop, the eye progresses; in LC, the eye cycles.

In the \texttt{for} version, fixations move progressively across multiple lines, following the sequential execution flow of the code. The reading pattern stabilizes over time, suggesting that participants build a coherent mental model. In the LC version, this progression does not occur. Fixations cycle between \texttt{listComprehension} and \texttt{input}, with repeated returns even after visiting \texttt{resultAssignment}.
This cyclical pattern helps explain the significant increase in AOI time for LC: participants spent more time inside the construct-specific AOI because they repeatedly revisited the comprehension line instead of progressing through separate execution steps.

\begin{figure}[]
\centering
\includegraphics[width=9cm]{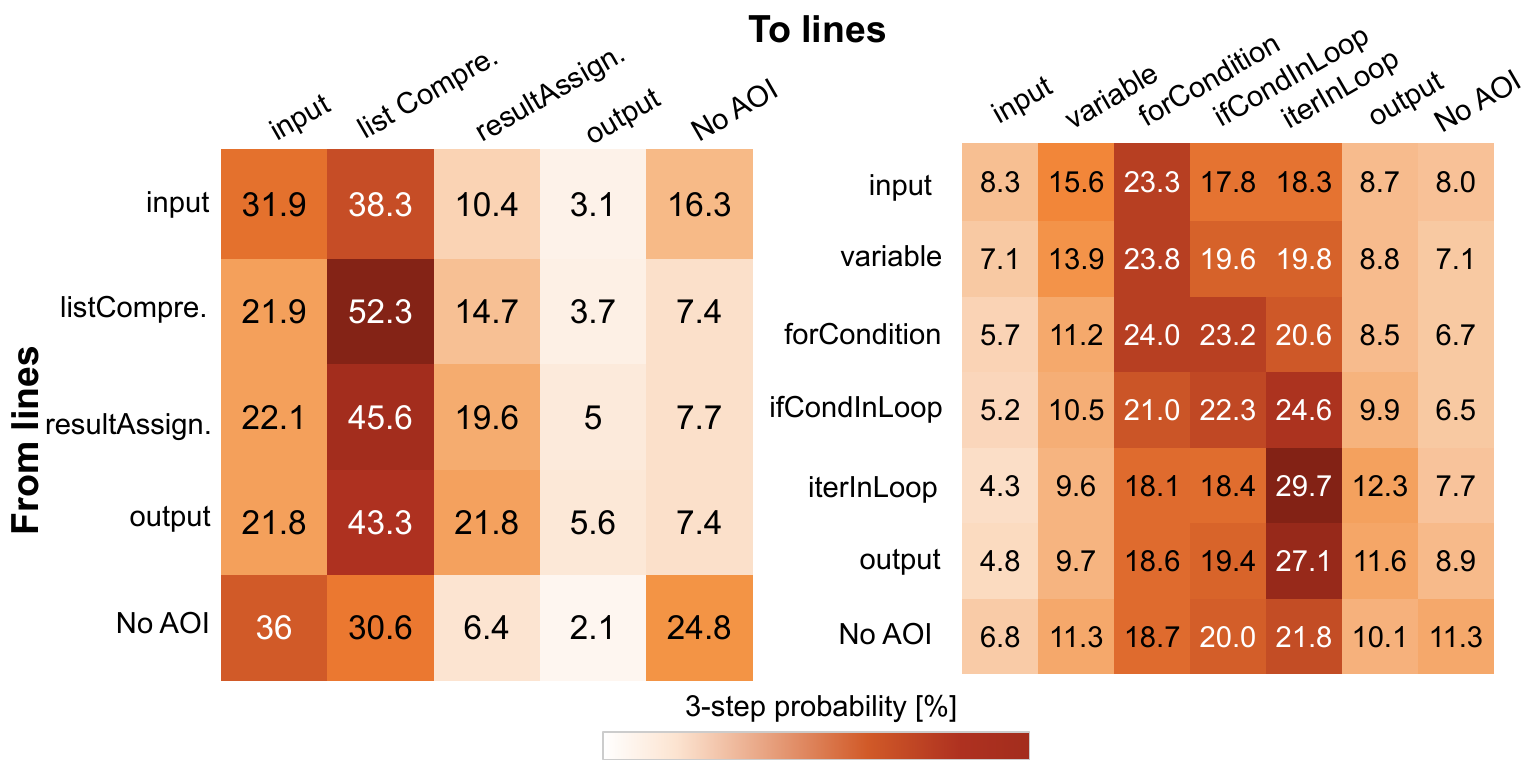}
\caption{Comparison of transition matrix of \texttt{for} and LC with 3-step transition probability.}
\label{fig:LCmatrix}
\end{figure}

The transition matrices corroborate this pattern quantitatively (Figure~\ref{fig:LCmatrix}). In these matrices, a 3-step transition probability indicates how likely the gaze is to move from one region of code to another within three successive fixation transitions, revealing short-range reading paths between code regions. In the LC version, the  \texttt{listComprehension} region shows a 3-step self-transition probability of 52.3\%, which is the highest value in the entire matrix, providing evidence that the eye is systematically trapped within the LC line. Even when participants reach \texttt{resultAssignment}, they return to \texttt{listComprehension} with 45.6\% probability, indicating that comprehension of the LC line is never fully resolved. In contrast, the \texttt{for} version shows no dominant self-transition, with values distributed across AOIs and a maximum of 29.7\%, reflecting a more exploratory and  progressive reading strategy. Therefore, the transition matrix provides a quantitative explanation for the AOI time and fixation-duration results: the LC line acts as a local processing bottleneck, forcing repeated short-range transitions within the same dense expression.

Participants' perceptions corroborate these findings: the \texttt{for} version was rated mostly as easy, while the LC version showed more neutral and difficult ratings. In interviews, students attributed difficulty with LC to its single-line density, while others noted that readability improves after understanding the syntax.  {The complete distributions of strategy and self-reported difficulty responses across all tasks, including LC, are available in the study's supplementary artifact package~\cite{anonymous2026supplementary}.} Together, the quantitative metrics, scarf plots, transition matrix, and interviews converge on the same interpretation: LC did not merely increase effort because it was shorter or unfamiliar; it concentrated iteration, filtering, and result construction into a dense local expression that novices repeatedly parsed{, in a pattern suggestive of token-by-token processing.} 
These findings suggest that LC should be introduced gradually, starting from the equivalent \texttt{for} loop, with the mapping made explicit.

\subsection{Perceived Difficulty}
\label{perceivedDifficulty}

Figure~\ref{fig: preferences Dissertacao} illustrates participants' perceptions of task difficulty across different structures.  Overall, the baseline \texttt{for} versions received more favorable ratings than the corresponding treatment versions. In the baseline group, most tasks were rated predominantly as very easy, easy, or neutral, whereas the treatment group shows a clearer shift toward neutral and difficult ratings.
For the \texttt{while} tasks, the perceived difficulty varied by task. In the \textit{Sum list} task, the \texttt{while} version still received many easy ratings, but it also had a larger difficult portion than its \texttt{for} counterpart. In the \textit{Multiples of three} task, the \texttt{while} version shifted strongly toward neutral ratings, suggesting that participants did not necessarily perceive it as very difficult, but found it less immediately easy than the \texttt{for} version. This pattern is consistent with the scarf plot analysis.

For recursion, the perceived-difficulty shift is more pronounced. In the \textit{Sum from one to n} task, the \texttt{for} version was mostly rated as easy or neutral, whereas the recursion version included a large proportion of difficult ratings. In the \textit{Exponentiation} task, the recursion version was also mostly rated as neutral or difficult. This supports the regression-graph evidence that participants had to mentally reconstruct the relationship between the base case, the recursive call, and the returned value.

For LC, the perception results also indicate higher difficulty than the corresponding \texttt{for} versions, but with variation across tasks. In the \textit{Quantity of 0 in list} task, the LC version still received many very easy or easy ratings, suggesting that some participants could understand the compact expression. However, the \textit{Count even numbers} LC task shifted more strongly toward neutral and difficult ratings. This variation suggests that LC difficulty is not only due to the construct itself, but also to how much logic is compressed into the comprehension expression. This interpretation is consistent with the transition matrix, where participants repeatedly returned to the \texttt{listComprehension} region of the code.

Taken together, the perceived-difficulty data support the main interpretation of the study: the treatment structures did not merely increase visual effort uniformly. Instead, each structure introduced a different source of comprehension demand: \texttt{while} tasks required explicit state tracking, recursion required call-flow reconstruction, and LC required parsing dense single-line expressions. Thus, participants' subjective perceptions converge with the quantitative eye-tracking metrics and the visual reading patterns. 

\section{Threats to Validity}
\label{sec:Threats to Validity}

Regarding \textbf{internal validity}, the experiment was conducted across three institutions, which may have introduced environmental variability. We mitigated this by standardizing lighting, temperature, and distractions across locations. Researcher presence was minimized during tasks to avoid influencing participants' behavior. Eye-tracking calibration is another relevant threat. The Tobii 4C, a low-cost eye tracker, has limited spatial accuracy compared with research-grade devices. {In cases of consistent vertical displacement between fixations and code lines, we applied a uniform \textit{y}-axis correction (10--70 pixels) to the affected task.} This correction was applied only when the displacement was systematic and visually evident, for example when fixations consistently fell on adjacent white space rather than on the code line being read. The correction procedure was discussed among the researchers and applied uniformly to the whole task, rather than selectively to individual fixations or AOIs. The Latin Square design minimized learning effects and task ordering bias. Individual differences in prior programming exposure and learning styles may have influenced results, though these were not the primary focus of this study.

Regarding \textbf{external validity}, short programs were used to fit the screen, consistent with similar eye tracking 
studies~\cite{da2021evaluating, da2023seeing, 
gopstein2017understanding}, but may limit generalization to larger codebases. In realistic maintenance and code-review scenarios, developers often inspect longer files, navigate across functions, use IDE support, and combine local reading with broader project knowledge. Our output-prediction tasks therefore capture a controlled form of program comprehension, but they do not fully represent maintenance, debugging, or refactoring activities in practice. {Our results are most directly applicable to localized comprehension scenarios involving short snippets, such as inspecting a small diff during code review or reasoning about a compact fragment before modifying it.} Participants were novice Python programmers, which limits generalization to experienced developers, though students have been shown to serve as valid proxies in certain contexts~\cite{salman2015students, falessi2018empirical}. {Nonetheless, we frame our claims to novice Python programmers, and plan to replicate our study with professional developers in future work.} 
The use of Portuguese identifiers and output-prediction tasks may further restrict generalization to other populations, programming contexts, and comprehension tasks. Portuguese identifiers reduced vocabulary barriers for our participants, but they may also affect how the results transfer to codebases that use English identifiers or different naming conventions. Additionally, our tasks involved a known and fixed number of iterations, which may limit the generalization to dynamic or unpredictable loop conditions.

Regarding \textbf{construct validity}, we combined traditional metrics (time, correctness) with eye-tracking data (fixation duration, fixation count, regressions), following established methodologies~\cite{da2021evaluating, da2023seeing, melo2017variability}. These measures capture visual effort but may not reflect the full complexity of cognitive processing. The comparison labeled ``All'' aggregates different treatment structures and tasks. This aggregation increases statistical power and provides an overall view of non-\texttt{for} structures, but it should not be interpreted as evidence that \texttt{while}, recursion, and LC impose the same kind of comprehension demand. Participants were not informed of all study details to reduce the risk of altered visual behavior during the tasks.

Regarding \textbf{conclusion validity}, the sample size of 32 participants may limit statistical power, and a larger sample would provide more robust results. This limitation is especially relevant for the pairwise comparisons, many of which showed large percentage differences but did not reach statistical significance. Therefore, non-significant pairwise results should be interpreted as descriptive trends rather than conclusive evidence of differences between individual constructs. Significance was set at 0.05. We also acknowledge that multiple metrics and comparisons increase the risk of chance findings; therefore, we interpret the results by considering the consistency of the quantitative metrics together with the eye-tracking visualizations and participants' qualitative feedback. The tasks, while representative of common introductory programming challenges, were limited in scope.

\vspace{-0.6cm}
\section{Related Work} 
\label{sec:Related Work} 

Several studies have compared iterative and recursive structures in introductory programming. Endres et al.~\cite{endres2021analysis} found that novices produce more errors with recursive and tail-recursive structures than with iterative ones in CS1 tasks. Sulov~\cite{sulov2016iteration} showed that novices prefer iteration over recursion, and McCauley et al.~\cite{mccauley2015teaching} identified building correct mental models as the central challenge in learning recursion. Haberman and Averbuch~\cite{haberman2002case} highlighted that novices specifically struggle with base cases, and Esteero et al.~\cite{esteero2018recursion} found that students who choose iteration in exam settings tend to perform better. None of these studies investigated the cognitive mechanisms underlying these differences; they relied mainly on performance metrics such as correctness and error rates. Thus, they show that recursion can be more difficult than iteration, but provide limited evidence about where this difficulty emerges during code reading.

List comprehensions have also been investigated in Python repositories. Prior work has analyzed their prevalence, complexity, evolution, fault-proneness, and performance characteristics~\cite{belias2022python,zampetti2022empirical,zid2024study}. They provide repository-level evidence about how LCs are used and maintained in practice, but they do not examine how developers visually process LC expressions during comprehension tasks. {In particular, Peng et al.~\cite{peng2021empirical} analyzed 35 popular Python projects comprising 25,059 files and approximately 4.3 million lines of code, measuring the usage of 22 language features. Their results show that \texttt{for} loops are among the five most frequently used features, accounting for 14.63\% of the reported occurrences.}

Eye tracking has been increasingly adopted to expose such mechanisms in code comprehension. Sharif and Maletic~\cite{sharif2010eye} first used it to compare identifier naming conventions, finding measurable differences in fixation patterns between camelCase and underscore styles. Jessup et al.~\cite{jessup2021using} compared expert and novice fixation patterns, finding that experts exhibit significantly more fixations than novices during comprehension tasks. Da Costa et al.~\cite{da2021evaluating} applied eye tracking to compare disciplined and undisciplined \#ifdef annotations in C, using fixation duration and regression count as visual effort metrics, which are also used in our study.

Da Costa et al.~\cite{da2023seeing} used eye tracking with 32 Python novices to study atoms of confusion with a similar experimental framework to ours. Da Costa et al.~\cite{da2026refactoring} studied how Java novices comprehend Extract and Inline Method refactorings using eye tracking. De Oliveira et al.~\cite{de2020atoms} similarly assessed confusion atoms through fixation and task completion metrics. Kather et al.~\cite{kather2021through} analyzed reading patterns to investigate how code familiarity influences mental model formation during comprehension tasks. Roberto et al.~\cite{pablo-sbes-2024} use eye tracking to investigate how Python code style influences developers’ visual attention and code comprehension. 

Our work extends this literature by applying eye tracking specifically to compare \texttt{for}, \texttt{while}, recursion, and LC within a single controlled experiment. Beyond fixation and regression metrics, we employ scarf plots and transition matrices to expose structure-specific reading patterns invisible to traditional performance metrics. Unlike prior work, our study compares four repetition structures within the same experimental design and connects aggregate eye-tracking metrics with structure-specific reading patterns.

\vspace{-0.4cm}
\section{Conclusions}
\label{sec:Conclusions}

We conducted a controlled eye-tracking experiment with 32 undergraduate students with prior Python experience to compare the comprehension of \texttt{for} loops, \texttt{while} loops, recursion, and list comprehensions across six Python tasks. Our results show that \texttt{for} loops were associated with the lowest visual effort across the analyzed metrics. Although not all pairwise comparisons reached statistical significance, the combined comparison between \texttt{for} loops and the other repetition structures was significant for all eye-tracking metrics. This additional effort was not uniform across constructs: \texttt{while} loops concentrated visual effort around counter initialization and update statements, recursion was associated with repeated transitions between the base case and the recursive call, and LCs produced horizontal regressions and repeated visits within the comprehension expression. These findings suggest that different repetition structures are associated with different comprehension demands rather than simply different levels of difficulty: 
the observed reading patterns suggest that \texttt{while} loops require explicit state tracking, recursion requires call-flow reconstruction, and LCs require dense local parsing.

These results have implications for software engineering practice. During code review, pair programming, onboarding, and maintenance, developers often need to understand code they did not write. In such contexts, choosing a repetition structure can affect readability and review effort, especially for less-experienced developers. Our results suggest that \texttt{for} loops may reduce comprehension overhead when iterations are simple and bounded. However, the practical implication is not that \texttt{while}, recursion, or LCs should be avoided. Instead, developers, reviewers, and educators should consider the specific source of comprehension effort introduced by each construct: explicit counter management, recursive base-case reasoning, or dense single-line transformations. 

As future work, we plan to replicate the study with a larger sample, more diverse tasks, and larger code snippets using scrolling-capable eye-tracking tools such as \textit{iTrace}~\cite{guarner2018itrac}. 
{We also aim to investigate more realistic maintenance tasks, IDE-supported reading settings, and more complex constructs such as nested list comprehensions and recursive implementations.} We further intend to investigate whether the observed visual-effort patterns generalize to experienced developers in professional settings. Finally, we aim to combine eye tracking with complementary methods, such as EEG~\cite{yeh2022identifying} and qualitative analysis based on grounded theory~\cite{strauss1998basics}, to deepen our understanding of how repetition structures affect code comprehension in realistic software development contexts.

\section*{Artifact Availability}
All study artifacts are available online~\cite{anonymous2026supplementary}.

\section*{Acknowledgments}
We want to thank the anonymous reviewers for their insightful suggestions. This work was partially supported by CNPq (306026/2026-0, 408040/2025-4, 403719/2024-0), CAPES (88887.313474/2026-00), FAPESQ-PB (268/2025).

\end{document}